\documentclass[12pt]{iopart}

\usepackage{graphicx}
\usepackage{braket}
\usepackage{array}
\usepackage{color}
\begin{document}

\title{Polariton blockade in the Jaynes--Cummings--Hubbard model with trapped ions}

\author{R Ohira$^1$, S Kume$^1$, H Takahashi$^2$ and K Toyoda$^3$}

\address{$^1$ Graduate School of Engineering Science, Osaka University, 1-3 Machikaneyama, Toyonaka, Osaka, Japan}
\address{$^2$ Experimental Quantum Information Physics Unit, Okinawa Institute of Science and Technology Graduate University, 1919-1 Tancha, Onna, Kunigami, Okinawa 904-0495, Japan}
\address{$^3$ Center for Quantum Information and Quantum Biology, Institute for Open and Transdisciplinary Research Initiatives, Osaka University, 1-3 Machikaneyama, Toyonaka, Osaka, Japan}
\ead{u696585a@ecs.osaka-u.ac.jp}

\vspace{10pt}
\begin{indented}
\item[]\today
\end{indented}

\begin{abstract}
We have experimentally observed the dynamics of a single polariton and two polaritons in a two-ion chain. By driving two trapped ions at a motional blue-sideband transition, we realize the anti-Jaynes--Cummings--Hubbard model. When a single polariton exists in a trapped-ion chain, the polariton hops between the ion sites. On the other hand, when there are single polaritons at each ion site, the hopping of the polaritons is suppressed because of the polariton--polariton interaction induced by the nonlinearity of the anti-Jaynes--Cummings interaction, thereby realizing the blockade of polariton hopping in the anti-Jaynes--Cummings--Hubbard model with trapped ions. Our work is a step towards the development of a trapped-ion based quantum simulator for strongly interacting polaritonic systems.
\end{abstract}

%
%
%
%

\section{Introduction}

The Jaynes--Cummings--Hubbard (JCH) model \cite{1,2} has been proposed to describe many-body physics in arrays of coupled optical cavities. Each cavity contains a two-level atom interacting with a photon, described by the Jaynes--Cummings (JC) model \cite{3}. In the JCH model, the atom in the cavity is dressed with photons, forming a polariton. The nonlinearity of the JC interaction introduces a polariton--polariton interaction, which leads to the advent of distinctive quantum phases of the polaritons.

Trapped ions are a promising candidate to physically implement the JCH model \cite{4,5}. In the JCH model with trapped ions, local phonons along a radial direction \cite{6,7,8,9,10} are used as bosonic particles instead of photons. The distance between the ions can be increased by weakening the confinement along the axial direction. In this situation, a picture of phonons localized to each ion site can be a good approximation for a short time period, and such `local phonons' exhibit hopping among different sites over a longer period. In addition to phonon hopping, illumination of an ion with a laser resonant with the motional sideband transition induces JC or anti-JC interactions \cite{11}. By driving every ion in a chain at a motional sideband transition, the JCH or anti-JCH model can be realized with trapped ions \cite{4,5}.

The concept of polaritons is important for understanding the quantum dynamics in the JCH model. In the JCH model with trapped ions, a polariton is expressed as the superposition of an internal excitation and a phonon. By tuning the power or frequency of the illuminating laser, it is possible to tune the strength of the polariton--polariton interactions. In this work, we have observed the dynamics of a single polariton and two polaritons in a two-ion chain. A single polariton freely hops between the two ion sites. However, in the presence of multiple polaritons, the polariton--polariton interaction induces a polariton blockade \cite{12,13}, where the hopping of the polaritons is suppressed.

The polariton blockade demonstrated in the present work can be considered as a phenomenon corresponding to the `photon blockade' in a cavity-QED system \cite{14}. In the photon-blockade experiment, the presence of an atom and the first photon in the cavity prevents the transmission of the second photon because of the anharmonicity in the eigenenergies of the coupled atom--cavity system. Similarly, in our experiment, the anharmonic energy levels at each ion site prevent the movement of polaritons. This can be recognized as the emergence of inter-particle interactions between the wave quanta (phonons or polaritons).

The polariton blockade presented in this work and the phonon blockade demonstrated in a previous work \cite{8} have a similarity in that the perturbed energy levels of local phonons at ion sites illuminated with sideband optical pulses play important roles. However, the mechanism of each blockade is different. In the phonon blockade experiment, a particular ion site is illuminated with a laser resonant with the sideband transition, preventing the local phonon from hopping into the illuminated site. This can be explained by a mismatch of energy levels between harmonic and anharmonic oscillators. On the other hand, in our experiment, two anharmonic oscillators with identical energy levels, which are coupled to each other, are realized. Nonetheless, the nonlinearity of the JC interactions introduces an effective polariton--polariton interaction, resulting in blocking of the excitation of the second polariton.

To precisely evaluate the polariton dynamics, we need to measure the quantum state in the polariton basis, which consists of both internal ($\ket{\downarrow}$ or $\ket{\uparrow}$) and phonon ($\ket{n=0, 1, 2\cdots}$) states. However, conventional state-dependent fluorescence detection does not allow the simultaneous evaluation of quantum states in both basis states. Despite this limitation, we can characterize the temporal dynamics of the ions by utilizing appropriate mapping sequences. Thereby, we have confirmed the presence of a polariton blockade by analyzing the experimental data along with numerical simulations.

\section{Theory}

\subsection{Anti-Jaynes--Cummings model}

Driving an ion at the red-sideband motional transition induces the JC interaction \cite{11}. The JC model can be described by the following Hamiltonian (with $\hbar$ set to 1, where $\hbar$ is Planck's constant $h$ divided by $2\pi$):
\begin{equation}
H_{\rm{JC}} = \omega\hat{a}^{\dagger}\hat{a}+(\omega+\Delta_{\rm JC})\hat{\sigma}^{+}\hat{\sigma}^{-}+g_{r}(\hat{a}\hat{\sigma}^{+}+\hat{a}^{\dagger}\hat{\sigma}^{-}),
\end{equation}
where $\hat{a}^{\dagger}$ and $\hat{a}$ are the creation and annihilation operators for phonons with energy $\omega$. The raising and lowering operators for the internal states are defined as $\hat{\sigma}^{+}=\ket{\uparrow}\bra{\downarrow}$ and $\hat{\sigma}^{-}=\ket{\downarrow}\bra{\uparrow}$. The transition frequency between $\ket{\downarrow}$ and $\ket{\uparrow}$ is $\omega_{0}$. $\Delta_{\rm JC}\equiv(\omega_{0}-\omega)-\omega_{\rm L},$ where $\omega_{\rm L}$ is the driving laser frequency. The last term corresponds to the red-sideband transition, where ${\it g_{r}}$ represents the JC coupling strength.

In the present experiment, we use the anti-JC interaction instead of the JC interaction, where the ions are illuminated with lasers resonant with a blue-sideband transition.\footnote{The blue-sideband transition is used in our experiment because the decay time of the contrast in blue-sideband Rabi oscillations is longer than that of the red-sideband Rabi oscillations.} The Hamiltonian for the anti-JC interaction is written as
\begin{equation}
H_{\rm{aJC}} = \omega\hat{a}^{\dagger}\hat{a}+(\omega+\Delta_{\rm aJC})\hat{\sigma}^{-}\hat{\sigma}^{+}+g_{b}(\hat{a}\hat{\sigma}^{-}+\hat{a}^{\dagger}\hat{\sigma}^{+}),
\end{equation}
where $\Delta_{\rm aJC}\equiv\omega_{\rm L}-(\omega_{0}+\omega)$. The last term corresponds to the blue-sideband transition, and ${\it g_{b}}$ represents the anti-JC coupling strength. Note that this anti-JC Hamiltonian is formally equivalent to the JC Hamiltonian given in equation\,(1) as they can be brought to the same form by interchanging the labeling of $\ket{\downarrow}$ and $\ket{\uparrow}$.

Then, at resonance, i.e. $\Delta_{\rm aJC}=0$, the eigenstates of $H_{\rm{aJC}}$ are
\begin{eqnarray}
\left\{
\begin{array}{l}
\ket{l=0}=\ket{\uparrow,0},\\
\ket{l^\pm} = \frac{1}{\sqrt{2}}(\ket{\uparrow,l}\pm\ket{\downarrow,l-1})\quad(l>0).
\end{array}
\right.
\end{eqnarray}
Here, $l$ represents the the number of polaritons. $\ket{\uparrow,l}\equiv\ket{\uparrow}\otimes\ket{n=l}$ and $\ket{\downarrow,l-1}\equiv\ket{\downarrow}\otimes\ket{n=l-1}$. The corresponding eigenenergies are given by
\begin{eqnarray}
\left\{
\begin{array}{l}
E_{0}=0\quad(l=0),\\
E_{\pm,l} = l\omega\pm\sqrt{l}g_{b}\quad(l>0).
\end{array}
\right.
\end{eqnarray}
The energy levels for the anti-JC eigenstates are shown in figure\,1(b).

\subsection{Anti-Jaynes--Cummings--Hubbard model with trapped ions}

The Hamiltonian for the anti-JCH model with $N$ ions with mass $m$ and charge $e$, in a frame rotating with frequency $\omega_r$ for harmonic confinement along a radial direction with the rotating wave approximation, is described as follows:
\begin{equation}
H_{\rm aJCH} = \sum_{i=\rm 1}^{N}\omega_{i}\hat{a}_{i}^{\dagger}\hat{a}_{i}+\sum_{i=\rm 1}^{N}\delta_{i}\hat{\sigma}_{i}^{-}\hat{\sigma}_{i}^{+}+g_{b}\sum_{i=\rm1}^{N}(\hat{a}_{i}\hat{\sigma_{i}}^{-}+\hat{a}_{i}^{\dagger}\hat{\sigma_{i}}^{+})+\sum_{i<j}^{N}\frac{\kappa_{ij}}{2}(\hat{a}_{i}\hat{a}_{j}^{\dagger}+\hat{a}_{j}^{\dagger}\hat{a}_{i}),
\end{equation}
where $\kappa_{ij}=e^{2}/4\pi\varepsilon_{0}\it{m}\it{d_{ij}}^{\rm 3}\omega_{r}$ and $\omega_{i}=-\sum_{i\neq j}^{N}\kappa_{ij}/{\rm 2}$ are the hopping rate and the position-dependent secular frequency shift of the $\it{i}$th ion, respectively. The first term represents the harmonic oscillator, where $\hat{a}_{i}^{\dagger}$ and $\hat{a}_{i}$ are the creation and annihilation operators for local phonons along the radial direction of the $\it{i}$th ion. The second term represents the energy of the $\it{i}$th ion, where the raising and lowering operators for the ${\it i}$th ion are defined as $\hat{\sigma}_{i}^{+}=\ket{\uparrow_{i}}\bra{\downarrow_{i}}$ and $\hat{\sigma}_{i}^{-}=\ket{\downarrow_{i}}\bra{\uparrow_{i}}$. $\delta_{i}$ is the detuning from the resonance of the blue-sideband transition for the $\it{i}$th ion. In the present work, we perform every experiment in the resonant condition, i.e., $\delta_{i}=0$. The third term represents the anti-JC interactions induced by a laser tuned to the blue-sideband transition. ${\it g_{b}}$ can be denoted as $\eta\Omega_{0}/{2}$, where $\eta$ is the Lamb--Dicke parameter and $\Omega_{0}$ is a value close to the Rabi frequency for the carrier transition. The last term describes hopping of phonons between different ion sites. $\kappa_{ij}$ is the hopping rate of local phonons between the $i$th and $j$th ion sites.

\subsection{Polaritons in trapped ions}

In the anti-JCH model with trapped ions, polaritons arise when the ions are illuminated with blue-sideband optical pulses. The energy levels for an ion incorporating both the internal and phonon states are shown in figure\,1(a).  

The anti-JCH model is characterized by the polariton number at each ion site. The polariton-number operator for the anti-JCH model at $i$th ion site is defined as $N_{i} = \hat{a}_{i}^{\dagger}\hat{a}_{i}+\hat{\sigma}_{i}^{-}\hat{\sigma}_{i}^{+}$. \footnote{Note that the definition of the internal-state excitation is different from the ordinary convention because of the anti-JC coupling: it gives an expectation value of 1 for the state $\ket{\downarrow_i}$.} In the present experiment, a polariton number of up to two is considered. Since the anti-JCH Hamiltonian commutes with the total polariton-number operator $N_{\rm{t}}$, i.e. $[H_{\rm{aJCH}}, N_{\rm{t}}] = [H_{\rm{aJCH}}, \sum_{i=\rm 1}^{N}N_{i}] = 0$, $N_{\rm{t}}$ is conserved in the anti-JCH system.

\begin{figure}[h]
\centering
  \includegraphics[width=14.0cm]{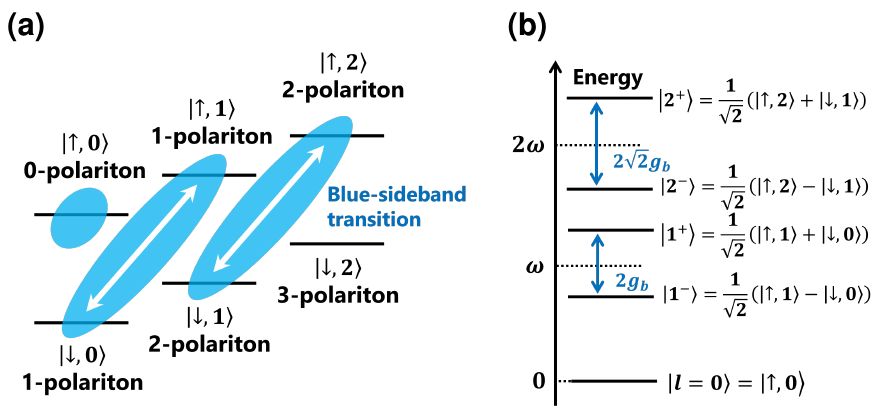}
\caption{\label{fig1} (a) Basis states relevant to polaritons in the anti-JC model with trapped ions. The unperturbed energy levels for an ion incorporating both the internal and phonon states are depicted. (b) Energy levels for the anti-JC eigenstates when $\Delta_{\rm aJC}=0$. Here the energy of $\ket{\downarrow,n}$ is identified with that of $\ket{\uparrow,n+1}$ by going into the frame rotating with their difference energy.}
\end{figure}

\section{Experimental results}

\subsection{Single-polariton hopping}

We first present single-polariton dynamics in a two-ion chain. The experimental setup is shown in figure\,2(a). Two $^{40}{\rm Ca}^{+}$ ions are trapped in a linear Paul trap. The frequencies for the harmonic confinement along the radial ($\it{x}$ and $\it{y}$) and axial ($\it{z}$) directions for the two ions are $(\omega_{x}, \omega_{y}, \omega_{z})/{\rm 2}\pi=$ (3.00, 2.81, 0.11) MHz. To encode a two-level system, we use the internal states $\ket{S_{1/2},m_{j}=-1/2}\equiv\ket{\downarrow}$ and $\ket{D_{5/2},m_{j}=-1/2}\equiv\ket{\uparrow}$ (see figure\,2(b)). The lifetime of the metastable state $D_{5/2}$ is $\sim1.2$ s. In addition, we use another Zeeman sublevel in the metastable state, $\ket{D_{5/2},m_{j}=-5/2}\equiv\ket{e_{0}}$, as an auxiliary state for shelving probability amplitudes in the detection sequence.

\begin{figure}[h]
\centering
  \includegraphics[width=14.0cm]{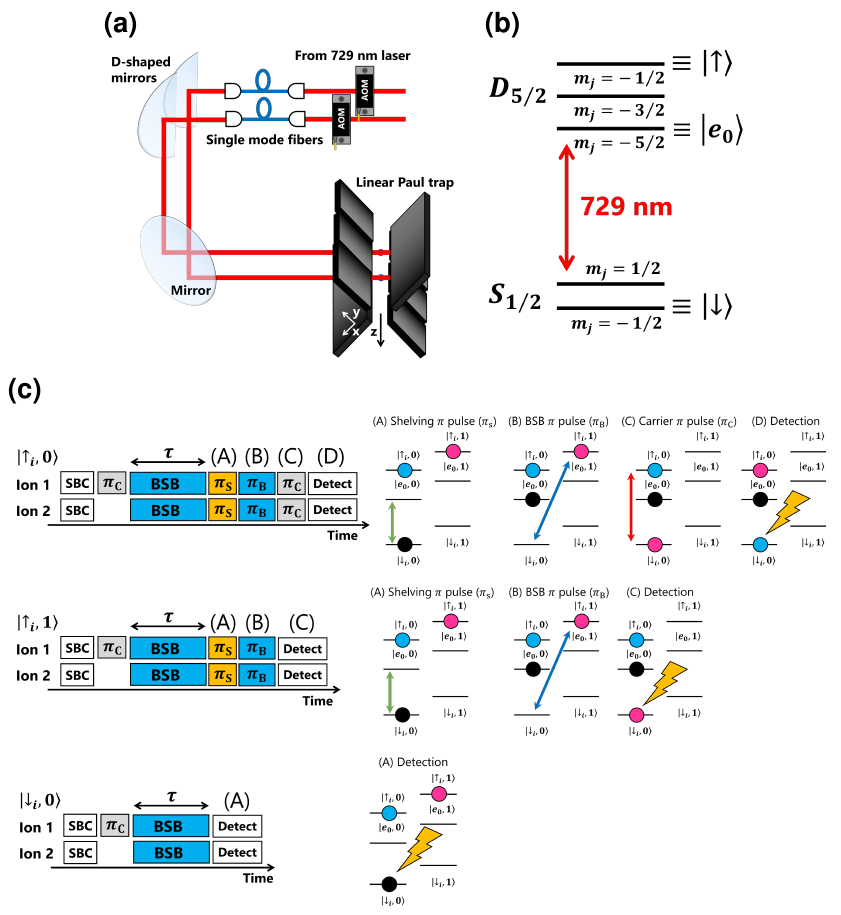}
\caption{\label{fig2} (a) Experimental setup: two $^{40}{\rm Ca}^{+}$ ions are trapped in a linear Paul trap. Two 729 nm beams are focused on each ion to manipulate their local phonon states. (b) Relevant energy diagram of $^{40}{\rm Ca}^{+}$ ion. Internal states $\ket{S_{1/2},m_{j}=-1/2}$ and $\ket{D_{5/2},m_{j}=-1/2}$ are used as $\ket{\downarrow}$ and $\ket{\uparrow}$, respectively. We also use the metastable state $\ket{D_{5/2},m_{j}=-5/2}\equiv\ket{e_{0}}$ in the experiments. (c) Experimental sequences to observe single-polariton hopping. To perform the measurement in three different basis states $\ket{\uparrow_{i},0}$, $\ket{\uparrow_{i},1}$, and $\ket{\downarrow_{i},0}$, one of the three different sequences is chosen and conducted. After sideband cooling (SBC), a carrier $\pi$ pulse ($\pi_{\rm C}$) is applied to Ion 1. Then, both ions are illuminated with lasers tuned to the blue-sideband (BSB) transition. After waiting for a duration $\tau$, a mapping pulse sequence is applied to measure the quantum states in each basis state. A shelving $\pi$ pulse ($\pi_{\rm S}$), which drives the transition $\ket{\downarrow_{i},n}\leftrightarrow\ket{{e_{0}}_{i},n}$, and a blue-sideband $\pi$ pulse ($\pi_{\rm B}$) are applied accordingly depending on the target state. Afterwards, the fluorescence detection is performed.}
\end{figure}

The experimental sequence is shown in figure\,2(c). In the experiments, three different transitions are used: carrier ($\ket{\downarrow_{i},n}\leftrightarrow\ket{\uparrow_{i},n}$), blue-sideband ($\ket{\downarrow_{i},n}\leftrightarrow\ket{\uparrow_{i},n+1}$), and ``shelving'' ($\ket{\downarrow_{i},n}\leftrightarrow\ket{{e_{0}}_{i},n}$). The experiment starts with Doppler cooling with lasers at 397 nm (${\it S_{\rm 1/2}}$--${\it P_{\rm 1/2}}$) and 866 nm (${\it D_{\rm 3/2}}$--${\it P_{\rm 1/2}}$) followed by resolved sideband cooling of the radial motional modes. A narrow quadrupole transition (729 nm, ${\it S_{\rm 1/2}}$--${\it D_{\rm 5/2}}$) is used for motional ground state cooling. In the experiment, we employ local phonons along the {\it y} direction. After the sideband cooling, the average quantum number for local phonons along the {\it y} direction is 0.04.

As shown in figure\,2(a), two individual 729 nm beams are focused on the ions for simultaneous addressing. For this simultaneous addressing, we use an optical system that is based on single-mode optical fibers. We prepare two individual beams at 729 nm with a polarization beam splitter and half waveplates. After passing through an acousto-optic modulator, each beam is coupled to a single-mode fiber for stabilizing the optical path. Then, the two beams are aligned with D-shaped mirrors. Compared with other optics such as half mirrors, D-shaped mirrors mitigate optical power losses so that fast Rabi flopping can be realized. By simply adding similar optical paths, we can address more individual ions simultaneously. The beam is focused to a size of $\sim3$ $\mu$m.

We prepare the ions in $\ket{\psi_{\rm Init}}=\ket{\psi_{1}}\otimes\ket{\psi_{2}}=\ket{\uparrow_{1},0}\otimes\ket{\downarrow_{2},0}=\ket{\uparrow_{1},0}\otimes(\ket{1^+}-\ket{1^-})/\sqrt{2}$ by applying a carrier $\pi$ pulse to Ion 1. Here, the polariton numbers for Ion 1 and Ion 2 in the anti-JCH model are 0 and 1, respectively. After preparing a single polariton, both ions are illuminated with a pulse tuned to the blue-sideband transition.

To experimentally observe the quantum dynamics of a single polariton, we need to measure the quantum states in three basis states, $\ket{\uparrow_{i},0}$, $\ket{\uparrow_{i},1}$, and $\ket{\downarrow_{i},1}$. Therefore, we employ three different sequences for three states, shown in figure\,2(c). By utilizing these mapping sequences, it is possible to map the probability amplitude of each state onto an internal state. For detection, 397 nm and 866 nm lasers are applied to collect the state-dependent fluorescence with a CCD camera.

\begin{figure*}[h]
\centering
  \includegraphics[width=16.0cm]{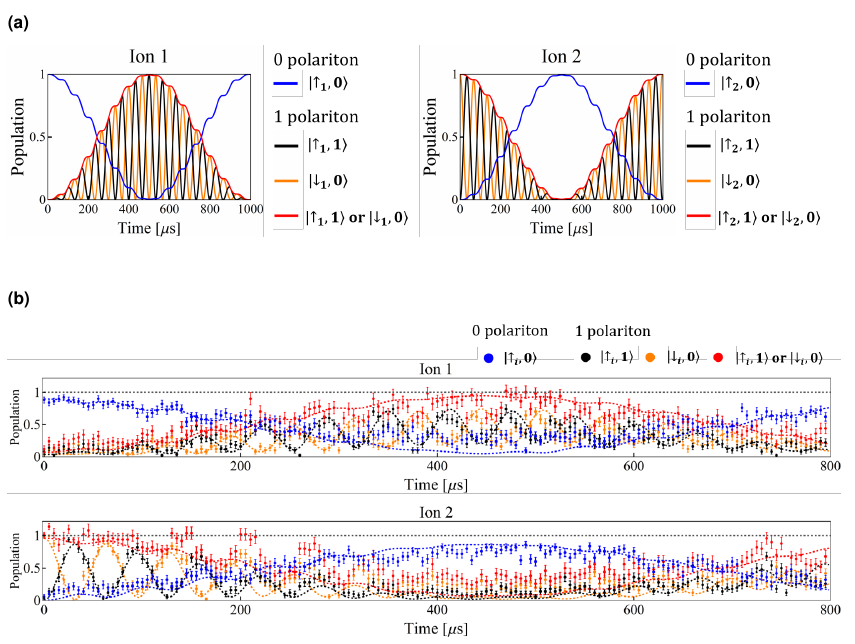}
\caption{\label{fig3} (a) Numerical simulation of single polariton hopping. Simulated populations in $\ket{\uparrow_{i},0}$ (blue), $\ket{\uparrow_{i},1}$ (black), and $\ket{\downarrow_{i},0}$ (orange), as well as the sum of the populations in $\ket{\uparrow_{i},1}$ and $\ket{\downarrow_{i},0}$ (red) are shown. The blue and red solid curves correspond to populations in 0- and 1-polariton manifolds, respectively. A single polariton hops between Ion 1 and Ion 2. (b) Results of the single-polariton hopping experiments of Ion 1 (top) and Ion 2 (bottom). Measured populations in $\ket{\uparrow_{i},0}$ (blue), $\ket{\uparrow_{i},1}$ (black), and $\ket{\downarrow_{i},0}$ (orange), as well as the sum of the populations in $\ket{\uparrow_{i},1}$ and $\ket{\downarrow_{i},0}$ (red), are shown. The blue and red data correspond to populations in 0- and 1-polariton manifolds, respectively. The dotted curves represent numerical calculations for each state or manifold. Each point is the average of 50 measurements. The gray dotted lines represent a population of 1.}
\end{figure*}

We first provide the results of numerical simulations. We employ the Lindblad master equation for the density matrix with parameters $(\kappa_{12}, 2g_{\rm b})/{\rm 2}\pi$ = (2, 15) kHz. The results are shown in figure\,3(a). As an initial state, the quantum states of the ions are prepared in $\ket{\psi_{\rm Init}}=\ket{\uparrow_{1},0}\otimes\ket{\downarrow_{2},0}$. For simplicity, we do not include any decoherence processes in the results of figure\,3(a). The population in the $l$-polariton manifold at the $i$th ion site ($i=1,2$) can be represented as 
\begin{eqnarray}
\left\{
\begin{array}{l}
\bra{\uparrow_{i},0}\rho_{i}\ket{\uparrow_{i},0}\quad(l=0),\\
\bra{\uparrow_{i},l}\rho_{i}\ket{\uparrow_{i},l}+\bra{\downarrow_{i},l-1}\rho_{i}\ket{\downarrow_{i},l-1}\quad(l>0),
\end{array}
\right.
\end{eqnarray}
where $\rho_{i}$ is the reduced density matrix for the {\it i}th ion. Accordingly, the population in the 0-polariton manifold can be identified as that in $\ket{\uparrow_{i},0}$, and the population in the 1-polariton manifold is the sum of those in $\ket{\uparrow_{i},1}$ and $\ket{\downarrow_{i},0}$. 

In figure\,3(a), the populations in the 0-polariton and 1-polariton manifolds are shown as blue and red solid curves, respectively. Additionally, the simulated populations in $\ket{\uparrow_{i},1}$ and $\ket{\downarrow_{i},0}$ are represented as black and orange solid curves, respectively. As seen in figure\,3(a), there are slow oscillations with a period of $\sim1$ ms in the blue or red curves, and fast oscillations with a period of $\sim70$ $\mu$s in the black or orange curves. The slow oscillations indicate that a single polariton hops between Ion 1 and Ion 2. The fast oscillations correspond to the blue-sideband transition between $\ket{\uparrow_{i},1}$ and $\ket{\downarrow_{i},0}$.

In figure\,3(b), the measured populations in $\ket{\uparrow_{i},0}$ (blue), $\ket{\uparrow_{i},1}$ (black), and $\ket{\downarrow_{i},0}$ (orange) are shown.\footnote{Note that the three populations shown here are the results of independent measurements and are not normalized with respect to their sum. We avoid normalizing them in order not to introduce additional systematic shifts of the populations. The errors of detecting a 1-polariton state may include systematic ones due to the fluctuation of the blue-sideband Rabi frequency, which are discussed later. Thus, normalizing the overall results may lead to the unwanted propagation of such systematic errors among the populations.} The blue and red data represent the measured populations in the 0-polariton and 1-polariton manifolds, respectively. Each point is the average of 50 measurements and the dotted curves are numerically calculated results incorporating the infidelity of the initial preparation and the dephasing of the Rabi oscillation at the carrier, blue-sideband, and shelving transitions. The blue-sideband Rabi frequency used in the numerical simulations is 15.7 kHz. The hopping rate $\kappa_{12}/{\rm 2}\pi$ is experimentally measured to be 2.12 kHz.

In figure\,3(b), the population in the 1-polariton manifold (red) at some data points exceeds 1. This is due to the fluctuation of the blue-sideband Rabi frequency in the measurements. During the experiment, we observe temporal changes of the ions' positions ($\sim0.6$ $\mu$m), which may lead to variations in the blue-sideband Rabi frequencies. The changes can be attributed to fluctuations in stray electric fields. Considering the beam waist ($\sim3$ $\mu$m), the corresponding reduction of the laser power experienced by the ions is estimated to be about 15\%. The population in the 1-polariton manifold (red) is deduced as being the sum of the populations in $\ket{\uparrow_{i},1}$ (black) and $\ket{\downarrow_{i},0}$ (orange). A fluctuation of the Rabi frequency between the measurements could cause errors in these populations such that the summation of the two becomes unphysical.  

\subsection{Polariton blockade}

We next present the results of the polariton blockade experiment. The polariton blockade studied in the present experiment is based on the nonlinearity of the anti-JC interactions. 

We consider a two-ion chain and assume that initially a single polariton exists at each ion site. The quantum state of the ions is expressed as $\ket{\psi}=\ket{1^\pm}\otimes\ket{1^\pm}\equiv\ket{1^\pm,1^\pm}$. Here, the $l$-polariton state is experessed as $\ket{l=0}=\ket{\uparrow,0}$ or $\ket{l^\pm} = (\ket{\uparrow,l}\pm\ket{\downarrow,l-1})/\sqrt{2}$ $(l>0)$. The polariton hopping induces the transitions $\ket{1^\pm,1^\pm}\leftrightarrow\ket{2^\pm,0}=\ket{2^\pm}\otimes\ket{l=0}$ and  $\ket{1^\pm,1^\pm}\leftrightarrow\ket{0,2^\pm}=\ket{l=0}\otimes\ket{2^\pm}$. According to equation (4), the energy of $\ket{k^\pm,l^\pm}$ ($k+l=L$) is
\begin{equation}
E_{k^\pm}+E_{l^\pm} = L\omega\pm(\sqrt{k}+\sqrt{l})g_{b}.
\end{equation}
The single-polariton hopping corresponds to the case of $L=1$. In this case, the total energies for the possible basis states ($\ket{1^\pm,0}=\ket{1^\pm}\otimes\ket{l=0}$, $\ket{0,1^\pm}=\ket{l=0}\otimes\ket{1^\pm}$) are equal if one of the eigenenergy  branches (+ or $-$) is selected. This energy is explicitly denoted as $E_{1^\pm}+E_{0}=\omega\pm g_{b}$. Therefore, there is no energy gap between the basis states in each branch, i.e. $\ket{1^+,0}$ and $\ket{0,1^+}$, or $\ket{1^-,0}$ and $\ket{0,1^-}$ (see figure\,4(a)(left) for the energy-level diagram).

\begin{figure}[h]
\centering
  \includegraphics[width=13.0cm]{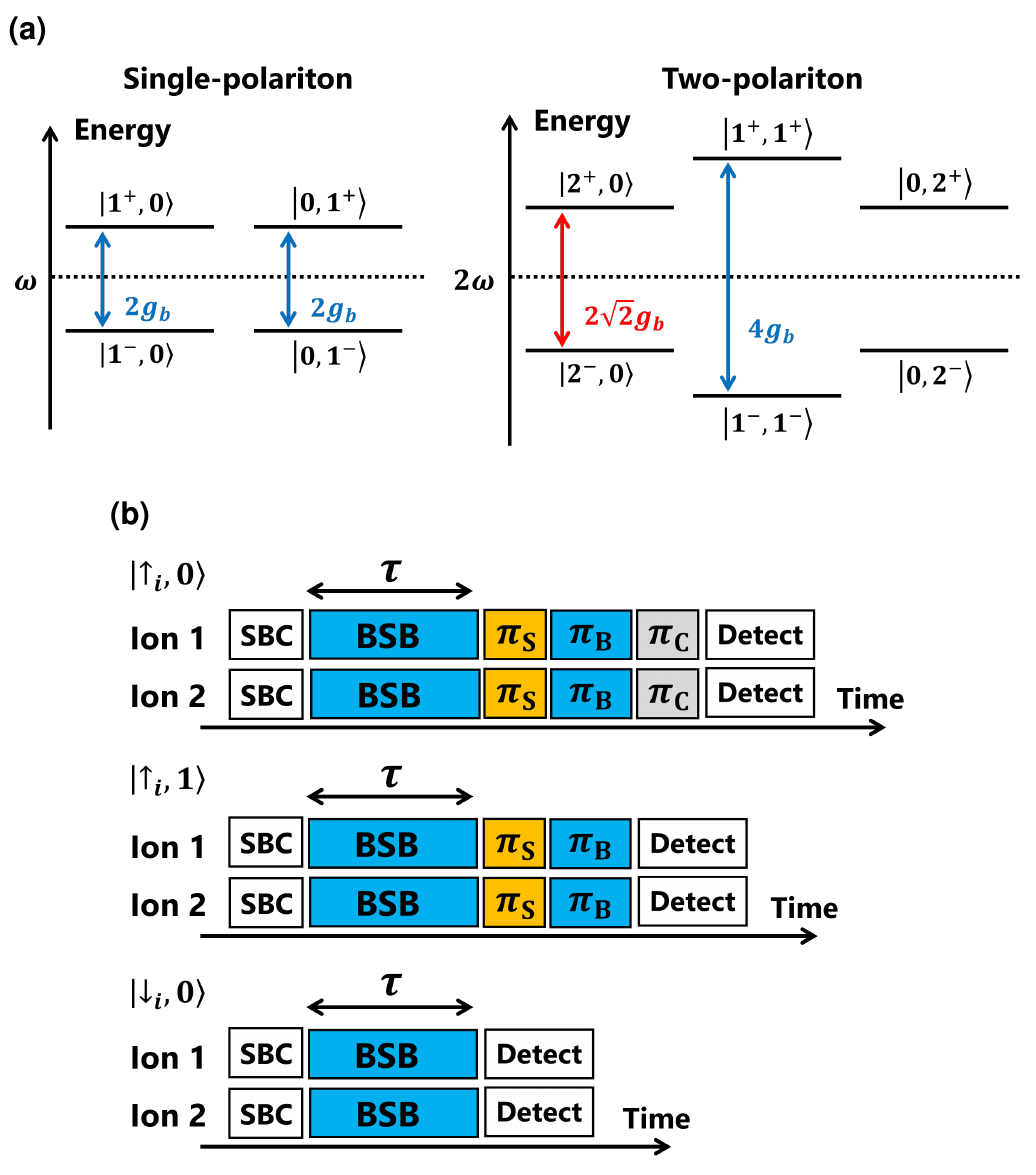}
\caption{\label{fig4} (a) Energy level diagram for a single-polariton (left) and two polaritons in two ions (right). (b) Experimental sequences for observing single-polariton blockades. One measurement sequence is chosen from the three different sequences, which are optimized for measurements in the basis states $\ket{\uparrow_{i},0}$, $\ket{\uparrow_{i},1}$, and $\ket{\downarrow_{i},0}$, respectively. Both ions are cooled to the motional ground states using sideband cooling (SBC). Then, both ions are illuminated with lasers detuned at the blue-sideband (BSB) transition. The illumination time of the BSB pulse $\tau$ is varied. After waiting for a duration $\tau$, a mapping pulse sequence is applied to measure the quantum states in each basis state. A shelving $\pi$ pulse ($\pi_{\rm S}$) and a blue-sideband $\pi$ pulse ($\pi_{\rm B}$) are employed depending on each state. Afterward, fluorescence detection is performed.}
\end{figure}

On the other hand, in the case of $L = 2$, there is an energy gap between $\ket{2^\pm,0}$ and $\ket{1^\pm,1^\pm}$ (or $\ket{1^\pm,1^\pm}$ and $\ket{0,2^\pm}$): $(E_{2^\pm}+E_{0})-2E_{1^\pm}=\mp(2-\sqrt{2})g_{b}\neq0$. This energy gap suppresses polariton hopping (see figure\,4(a)(right) for the energy-level diagram).
Note that a polariton blockade can be applied to the ${\it l}$-polariton case. The energy gap between $\ket{l^\pm}$ and $\ket{(l-1)^\pm}$ can be calculated as
\begin{equation}
\delta E_{l} = \omega\pm(\sqrt{l}-\sqrt{l-1})g_{b}.
\end{equation}
However, this indicates that as the polariton number $l$ grows, it becomes harder to block the transition $\ket{(l-1)^\pm}\leftrightarrow\ket{l^\pm}$. 

The experimental sequence is shown in figure\,4(b). After sideband cooling, the quantum state of the ions is $\ket{\psi_{\rm Init}}=\ket{\downarrow_{1},0}\otimes\ket{\downarrow_{2},0}$. Since $\ket{\downarrow_{i},0}$ is a superposition of the 1-polariton states, $\ket{1^+}$ and $\ket{1^-}$, no additional pulses are required for state preparation. Then, the ions are illuminated with blue-sideband pulse. For detection, we also employ the same pulse sequences used in the single-polariton experiment.

We first show the numerically simulated results for the quantum dynamics of two polaritons in a two-ion chain in figure\,5(a-c). Values for the parameters of $(\kappa_{12}, 2g_{\rm b})/{\rm 2}\pi$ = (2, 15) kHz are used for the numerical calculations. As an initial state, the quantum states of the ions are prepared in $\ket{\psi_{\rm Init}}=\ket{\downarrow_{1},0}\otimes\ket{\downarrow_{2},0}$. In this simulation, we do not include any decoherence processes. Note that in this case the dynamics of the two ions, as well as their initial states, are expected to be identical to each other due to the symmetry with respect to the exchange of the ions. Therefore, we only show the state populations for one of the ions.

\begin{figure*}[h]
\centering
  \includegraphics[width=16.0cm]{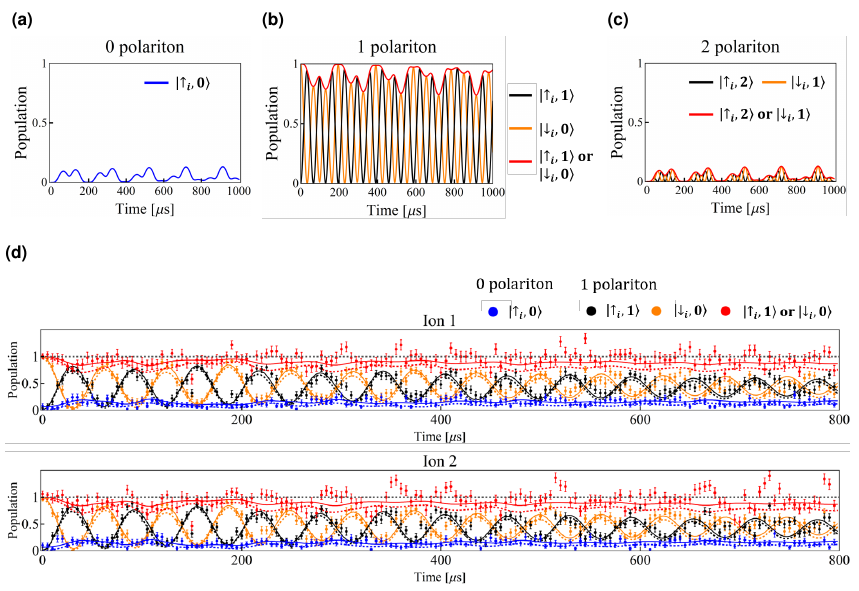}
\caption{\label{fig5} (a--c) Numerical simulation of a polariton blockade. (a) Simulated populations in $\ket{\uparrow_{i},0}$ (blue). (b) Simulated populations in $\ket{\uparrow_{i},1}$ (black), $\ket{\downarrow_{i},0}$ (orange), as well as their sum (red). (c) Simulated populations in $\ket{\uparrow_{i},2}$ (black) and $\ket{\downarrow_{i},1}$ (orange), as well as their sum (red). (d) Results of the single-polariton blockade experiments of Ion1 (top) and Ion 2 (bottom). Measured populations $\ket{\uparrow_{i},0}$ (blue), $\ket{\uparrow_{i},1}$ (black), and $\ket{\downarrow_{i},0}$ (orange), as well as the sum of $\ket{\uparrow_{i},1}$ and $\ket{\downarrow_{i},0}$ (red) are shown. Each point is an average of 50 measurements. The dotted curves represent numerically calculated populations. The solid curves correspond to the numerically calculated populations expected for each of the three actual measurement sequences in figure\,4(b). The gray dotted lines represent a population of 1.}
\end{figure*}

Figure\,5(a) shows the population in $\ket{\uparrow_i,0}$, or equivalently that in the 0-polariton manifold, as a blue solid curve. In figure\,5(b), the black and orange solid curves represent the simulated populations in $\ket{\uparrow_{i},1}$ and $\ket{\downarrow_{i},0}$, respectively, and the red solid curve corresponds to the population in the 1-polariton manifold. We also show the populations in the 2-polariton states in figure\,5(c). The black and orange solid curves represent the populations in $\ket{\uparrow_{i},2}$ and $\ket{\downarrow_{i},1}$, respectively. The red solid curves corresponds to the 2-polariton population ($\bra{\uparrow_{i},2}\rho_{i}\ket{\uparrow_{i},2}+\bra{\downarrow_{i},1}\rho_{i}\ket{\downarrow_{i},1}$). As can be seen in figure\,5(a--c), the population in the 1-polariton manifold at each ion site remains high while the population in 0-polariton and 2-polariton manifolds remains negligible.

In this experiment, for the full determination of the polariton states, we need to measure the 0-, 1-, and 2-polariton states. This requires measurements in five different basis states: $\ket{\uparrow_{i},0}$ for the 0-polariton state, $\{\ket{\uparrow_{i},1},\ket{\downarrow_{i},0}\}$ as the 1-polariton states, and $\{\ket{\uparrow_{i},2},\ket{\downarrow_{i},1}\}$ as the 2-polariton states. Fully determining the polariton state using state-dependent fluorescence detection requires many auxiliary levels and extensive pulse sequences similar to those used in phonon-number-resolving detection \cite{9} (see Discussion for details). However, in the present experimental setup, the decoherence of the relevant transitions required for the phonon-number-resolving detection is not negligible, and the overall results may result in the lower contrasts.

However, the measurement scheme can be simplified if we can assume that the system stays in the subspace with two polaritons in total. In this case, the population in the 0-polariton manifold at either of the two ion sites is equal to that in the 2-polariton manifold at the other ion site. Therefore, a simplified measurement scheme similar to that used in the observation of single-polariton hopping, where only $\ket{\uparrow_i,0}$, $\ket{\uparrow_i,1}$, and $\ket{\downarrow_i,0}$ are determined, would give the full information of the system with respect to the polariton number at each ion site.

Of course, the assumption given above is not exactly satisfied in reality, and we should consider deviations from it. Deviations toward smaller numbers of polaritons (i.e., zero or one polariton in total) can be covered by the simplified scheme given above, while those toward larger numbers of polaritons (i.e., three or more polaritons in total) should be considered separately. In our estimation, the maximum population that the system has residing in the subspaces with three polaritons or more in total is $\sim0.084$. This residual population can be estimated by considering imperfect preparation and population leakages due to the motional heating. The predominant cause of imperfect preparation is residual motional excited-state populations after sideband cooling of the relevant radial mode. The finite average quantum number is 0.04, indicating that residual populations of 0.08 exist for three polaritons or more in total. Leakage due to heating of the radial modes can be evaluated from independent measurements of the system's heating rate. Our previous study gives 5 quanta/s for a four-ion chain \cite{10}, and typically a smaller value is expected for the two ions used here. If we choose to use the same heating rate and a maximum total duration of the time sequence of $\sim840$ $\mu$s, we can obtain an upper bound of 0.004 for the current study. As a result, the maximum total population that the system has residing in the subspace with three polaritons or more is estimated to be $\sim0.084$. Although this amount is not negligible, we can interpret the qualitative behavior of the polaritonic system in the presence of such an imperfection.

Figure\,5(d) shows the experimentally observed populations in $\ket{\uparrow_{i},0}$ (blue), $\ket{\uparrow_{i},1}$ (black), $\ket{\downarrow_{i},0}$ (orange)\footnote{For the same reason as in the case of single-polariton hopping, these three populations are not normalized with respect to their sum.}, and the 1-polariton manifold (red). Each point is an average of 50 measurements. The dotted curves are numerically simulated populations in each state or manifold. The infidelity of the initial preparation and the dephasing of the Rabi oscillation at the carrier, blue-sideband, and shelving transitions are included in the simulation. The blue-sideband Rabi frequency used in the numerical simulations for each quantum state is 15.7 kHz. The experimental data show good agreement with the simulated data.

In addition to the deviations of the total number of polaritons mentioned above, there are two other major errors. The first is the fluctuation of the blue-sideband Rabi frequency, which results in some data points for the population in the 1-polariton manifold (red), exceeding 1. As discussed in the single-polariton hopping experiment, this can be attributed to a temporal shift in the position of the ions. The second derives from miscounting the state populations due to an incomplete experimental sequence. The experimental sequence used in this work is optimized for evaluating 0- and 1-polariton state. However, populations in higher polariton manifolds are also partially mapped to $\ket{\downarrow_{i}}$. These residual populations can cause measurement errors. To verify this error, we numerically calculate the populations expected for each of the three actual measurement sequences in figure\,4(b). The simulated populations are shown as solid curves in figure\,5(d). The simulated results incorporating the actual measurement sequences show better agreement with the experimental results, while the remaining discrepancies can be ascribed to the other error sources discussed above. The concentration of the population to the 1-polariton manifolds confirmed here indicates the occurrence of polariton blockades.

\section{Discussion}

Although the mapping sequences used in the present experiments are not exhaustive for detecting multiple polaritons, we speculate that we can characterize the polariton states using sufficient mapping sequences. This is because the populations in 0-polariton and 2-polariton manifolds are expected to be small and, therefore, we can distinguish the 3-polariton states with such non-exhaustive measurements. Nonetheless, we need a detection method to precisely evaluate the $n$-polariton state to explore the different polariton dynamics in the future, especially in regions where hopping is dominant compared with the polariton--polariton interaction. In this parameter region, the polariton hopping results in the superposition of different polariton states. 

One of the possibilities is the extended scheme of the phonon-number-resolving detection \cite{9}. Here, we show the possible detection scheme for observing the 2-polariton state (figure\,6). For implementing the proposed detection scheme, four long lived states are required, and in the case of $^{40}\rm Ca^{+}$ ion, the zeeman sublevels of $D_{5/2}$ state are used. Here, each state is labelled as $\ket{D_{5/2},m_{j}=-5/2}\equiv\ket{e_{0}}, \ket{D_{5/2},m_{j}=-3/2}\equiv\ket{e_{1}}, \ket{D_{5/2},m_{j}=1/2}\equiv\ket{e_{2}}$, and $\ket{D_{5/2},m_{j}=3/2}\equiv\ket{e_{3}}.$ The detection process is as follows. (1) A shelving $\pi$ pulse is applied to transfer the probability amplitude of $\ket{\downarrow_i}$ to $\ket{e_{3i}}$. (2) Either a composite-pulse sequence or adiabatic passage over the blue-sideband transition is applied to realize the uniform transfer of the probability amplitude between the transitions $\ket{\downarrow_{i},0}\leftrightarrow\ket{\uparrow_{i},1}$ and $\ket{\downarrow_{i},1}\leftrightarrow\ket{\uparrow_{i},2}$. (3) We, then, flip the internal state ($\ket{\downarrow_{i}}\leftrightarrow\ket{\uparrow_{i}}$) by applying a carrier $\pi$ pulse. (4) The probability amplitude of $\ket{\downarrow_{i},0}$ is transferred to $\ket{e_{0i},0}$ by a shelving $\pi$ pulse. (5) The probability amplitude of $\ket{\uparrow_{i},1}$ is transferred to $\ket{\downarrow_{i},0}$ by a blue-sideband $\pi$ pulse, then, is subsequently transferred to $\ket{e_{1i},0}$ by a shelving $\pi$ pulse. (6) Likewise, the probability amplitude of $\ket{\uparrow_{i},0}$ is mapped onto the long-lived auxiriary state $\ket{e_{2i},0}$ by sequential carrier and shelving $\pi$ pulses. (7) The probability amplitude of $\ket{e_{3i}}$ is transferred back to $\ket{\downarrow_{i}}$. (8) Finally, a red-sideband $\pi$ pulse maps the probability amplitude of $\ket{\downarrow_{i},1}$ onto $\ket{\uparrow_{i},0}$. After the mapping sequence, the all probability amplitudes of 0-, 1-, and 2-polariton basis states are stored in the motional ground state, i.e. $\ket{\uparrow_{i},0}\rightarrow\ket{e_{0i},0}$, $\ket{\uparrow_{i},1}\rightarrow\ket{e_{2i},0}$, $\ket{\downarrow_{i},0}\rightarrow\ket{\downarrow_{i},0}$, $\ket{\uparrow_{i},2}\rightarrow\ket{e_{1i},0}$, and $\ket{\downarrow_{i},1}\rightarrow\ket{\uparrow_{i},0}$. These mapped states can be detected by applying a sequential detection pulse as performed in the previous study \cite{9}. In the present experimental setup, the decoherence of the relevant transitions required for implementing this measurement is not negligible; therefore, we employ another detection scheme.

\begin{figure*}[h]
\centering
  \includegraphics[width=15.0cm]{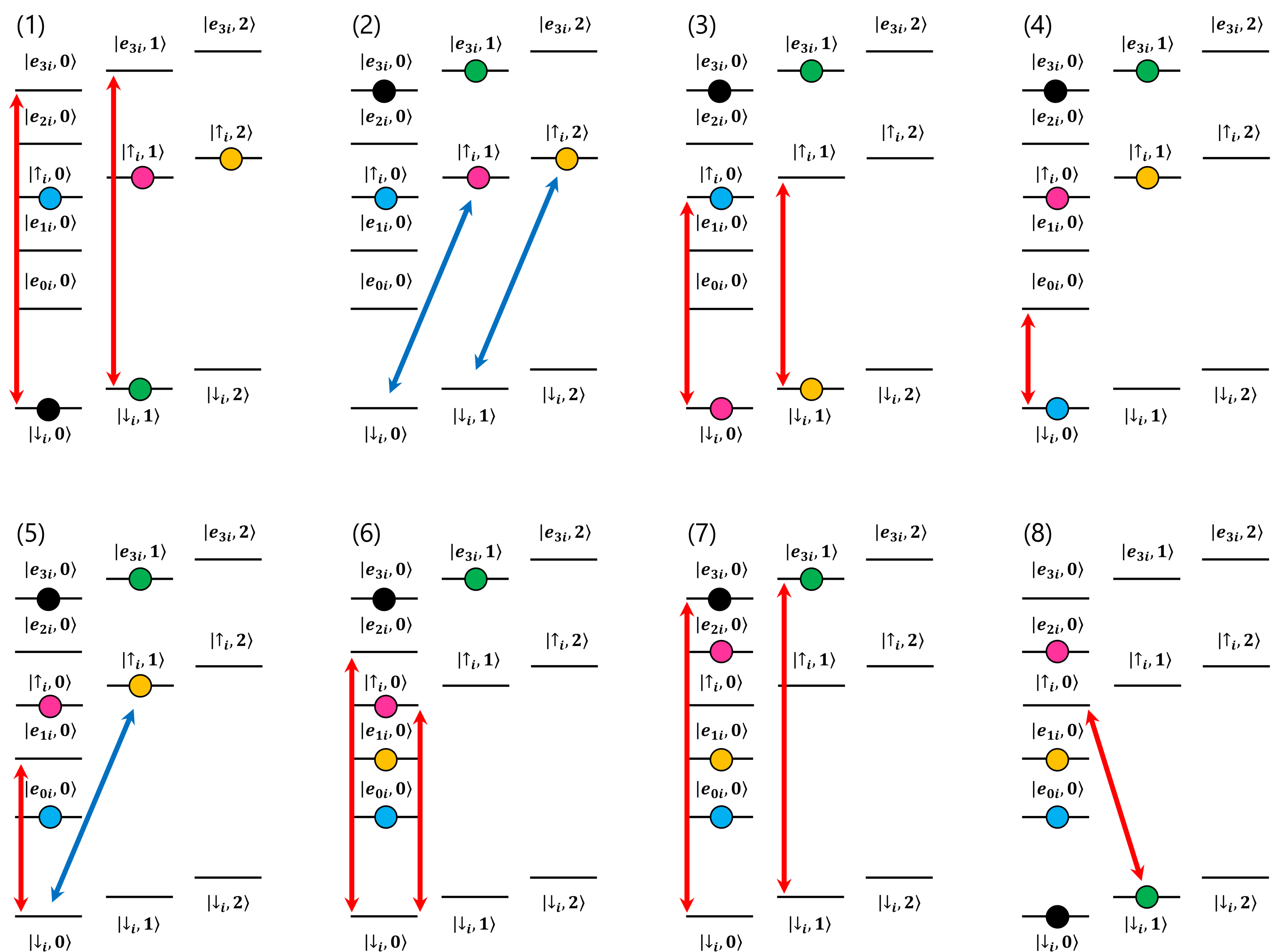}
\caption{\label{fig6} Complete mapping sequence for detecting the 2-polariton state. The probability amplitude of each polariton basis state is mapped onto a Zeeman sublevel of the metastable state with the vibrational motion in the ground state.}
\end{figure*}

The polariton--polariton interaction may play important roles in various physical phenomena expected to obey the JCH (anti-JCH) model. For instance, for 2-polariton quantum dynamics, it has been theoretically predicted that a 2-polariton bound state is expected to be observed \cite{15}, which is analogous to the second-order tunneling of a bound two-boson realized with cold atoms \cite{16,17}. By providing site-dependent polariton--polariton interactions, so that the polaritons feel a tilted potential, it may be possible to observe the Bloch oscillation of polaritons, which has also been extensively studied in cold atoms trapped in optical lattices \cite{18}.

\section{Conclusions}
In conclusion, we have observed the 1-polariton and 2-polariton dynamics in the anti-JCH model with two ions. A single polariton hops between the ion sites, whereas the polariton--polariton interaction suppresses the hopping in the presence of a single polariton at each ion site.

\section*{Acknowledgments}
This work was supported by MEXT Quantum Leap Flagship Program (MEXT Q-LEAP) Grant Number JPMXS0118067477.

\end{document}